\documentclass[preprint,authoryear,12pt]{elsarticle}

\usepackage{graphicx}
\usepackage{amssymb}
\usepackage{bm}
\usepackage{lscape}

\newcommand{\apj}{Astrophysical Journal}
\newcommand{\mnras}{Monthly Notices of the Royal Astronomical Society}
\newcommand{\pasj}{Publications of the Astronomical Society of Japan}
\newcommand{\newa}{New Astronomy}

\journal{Astronomy and Computing}

\begin{document}

\begin{frontmatter}



\title{Few-Body Modes of Binary Formation in Core Collapse}


\author[RIKEN,UTsukuba,UAizu]{Ataru Tanikawa\corref{cor1}}
\author[Edinburgh]{Douglas C. Heggie} \author[Princeton,ELSI]{Piet
  Hut} \author[RIKEN,ELSI]{Junichiro Makino}

\address[RIKEN]{RIKEN Advanced Institute for Computational Science,
  7--1--26, Minatojima-minami-machi, Chuo-ku, Kobe, Hyogo, 650--0047,
  Japan} \address[UTsukuba]{Center for Computational Science,
  University of Tsukuba, 1--1--1, Tennodai, Tsukuba, Ibaraki
  305--8577, Japan} \address[UAizu]{School of Computer Science and
  Engineering, University of Aizu, Tsuruga, Ikki-machi,
  Aizu-Wakamatsu, Fukushima, 965--8580, Japan}
\address[Edinburgh]{School of Mathematics and Maxwell Institute for
  Mathematical Sciences, University of Edinburgh, King's Buildings,
  Edinburgh EH9 3JZ} \address[Princeton]{Institute for Advanced Study,
  Princeton, NJ 08540, USA} \address[ELSI]{ Earth-Life Science
  Institute, Tokyo Institute of Technology, 2--12--1 Ookayama, Meguro,
  Tokyo 152-8551, Japan}

\cortext[cor1]{e-mail: ataru.tanikawa@riken.jp, TEL: +81-78-940-5865}

\begin{abstract}
  At the moment of deepest core collapse, a star cluster core contains
  less than ten stars.  This small number makes the traditional
  treatment of hard binary formation, assuming a homogeneous
  background density, suspect.  In a previous paper, we have found
  that indeed the conventional wisdom of binary formation, based on
  three-body encounters, is incorrect.  Here we refine that insight,
  by further dissecting the subsequent steps leading to hard binary
  formation. For this purpose, we add some analysis tools in order to
  make the study less subjective. We find that the conventional
  treatment does remain valid for direct three-body scattering, but
  fails for resonant three-body scattering.  Especially democratic
  resonance scattering, which forms an important part of the
  analytical theory of three-body binary formation, takes too much
  space and time to be approximated as being isolated, in the context
  of a cluster core around core collapse.  We conclude that, while
  three-body encounters can be analytically approximated as isolated,
  subsequent strong perturbations typically occur whenever those
  encounters give rise to democratic resonances.  We present
  analytical estimates postdicting our numerical results.  If we only
  had been a bit more clever, we could have predicted this qualitative
  behaviour.
\end{abstract}


\end{frontmatter}

% \linenumbers


\section{Introduction}

In our previous paper, \cite{Tanikawa11}, hereafter referred to as
Paper I, we started to investigate in detail the formation mechanism
of the first hard binary during core collapse of a dense star cluster.
While many studies have appeared that have focused on the macroscopic
aspects of core collapse, during the last fifty years, to the best of
our knowledge our paper was the first one to address the microscopic
aspects, including the actual reaction network of the stellar
encounters that gave rise to the formation of a hard binary.

In that study, we encountered two surprising deviations from what had
become accepted as the standard picture of binary formation in core
collapse.  First, in many cases more than three bodies are directly
and simultaneously involved in the production of the first hard
binary.  Second, we concluded that the core at deepest collapse was
smaller than expected before, typically containing half a dozen stars
or less.

In contrast, in the standard picture developed in the nineteen
eighties, it was first assumed that the formation of hard binaries was
essentially a three-body process, whose rate could be estimated
assuming the typical density and velocity dispersion in the core.
Second, it was concluded that the core would bounce around the time
its membership had dropped to a few dozen stars. In Paper I, we cited
papers by \cite{Goodman84,Goodman87}.  An additional reference is
\cite{Hut85}, where analytical arguments were used to predict that
three-body binary formation would reverse core collapse when the core
shrank to contain of order 100 stars (80 in their section IVa, and 150
in their section IVbii).  They also quoted simulations by
\cite{McMillan84} which showed core collapse to be reversed when the
core contained 25 stars.

The two main flaws in the traditional picture are related.  Given that
the fluctuations in thermodynamic properties in a group of only a few
stars are far larger than in a group of, say, thirty stars, the
concept of a homogeneous temperature (or velocity dispersion) in the
core is no longer valid for such a small core. Also, in a core
containing only, say, five stars it is quite likely that all five are
involved in the formation of a hard binary, with possibly some of the
stars just outside the core also making a strong presence felt during
a pass through the core.

Encouraged by the fact that the standard story of hard binary
formation needed to be corrected on at least these two quite
fundamental points, we continued our investigation, focusing in on
only one of the many runs reported in Paper I, in an attempt to get
further to the bottom of what is actually happening during core
collapse in microscopic detail.  Not wanting to introduce any bias, we
decided to simply take the very first case described in Paper I.

The central new technique, introduced in Paper I, was to plot all
pair-wise distances for all stars in the core, as a function of time,
during a short period of time just before hard binary formation.
Using this technique, and interpreting the results by eye, was only
feasible given the very small number of stars in the core.  Together
with visual interactive inspection of the 3-D orbits of the stars in
the core, our new technique allowed a determination of roughly how
many stars were involved at each time during the successive stages
leading to the formation of the first hard binary.

In this paper, we move beyond the detection of the new physics
reported in Paper I, i.e. many-body binary formation, in order to
perform a more detailed and quantitative analysis of this
binary-formation process. In particular, we devote efforts to finding
stars involved with the binary formation more objectively, and to
reveal what kind of subsystems these stars construct, and how these
stars and subsystems interact with each other. For this purpose, we
introduce two other new techniques. The first one is the use of work
functions, and the second one is a form of subcluster analysis. In
addition, we have employed a better interactive visualisation tool, in
the form of an {\tt open-GL} program. These tools will be useful to
make clear binary formation in more realistic and complicated $N$-body
simulations in which stars have different masses and experience
internal evolution.

In the process of applying these new tools, we again found new
physics: while the main conclusions of Paper I hold, we now understand
in more detail exactly why they hold. The main reason is the presence
of democratic resonance interactions, a concept introduced by
\cite{Hut82}, which is a kind of encounter between a hard binary and a
single intruder in which a third body is temporarily bound to the
binary, in such a way that the subsequent motion cannot be described
as a hierarchical triple system. In contrast, we found that the
traditional perturbative treatment is in fact satisfactory for direct
three-body interactions.  It is only because democratic resonance
interactions last long and take up a large fraction of the space in
the core that they will typically undergo strong encounters with other
stars before a democratic resonance is finished.

This new paper has two main aims: first, to illustrate the new
diagnostic tools and their uses (with a long-term goal of making this
kind of analysis more streamlined and automatic), and, second, to
explain our developed understanding of the new dynamical processes
which we are exploring.  This main finding is described in more detail
in the discussion and conclusion sections below.  The next two
sections focus on a summary of what we found out about the first run
in Paper I; and on what we learned in our new analysis in this paper,
respectively.  The section after that extends the analysis of Paper I
to earlier times, where interesting processes were already happening
that had not been flagged in Paper I. The paper finishes with a
section of theoretical discussion, and then a summary of our
conclusions and some outlook.

\section{Summary of the first run in Paper I}\label{sec:summaryI}

The analysis in Paper I consisted mostly of inspection-by-eye, which
sufficed to find the new physical phenomena, mainly the fact that more
than three stars were involved in most instances of hard binary
formation.  Here follows a brief summary of the first run of Paper I,
the 1024-body run with seed~1.

We performed an $N$-body simulation of a star cluster. At the initial
time, it had $1024$ stars with equal masses, and a stellar
distribution given by the Plummer model. For the $N$-body simulation,
we have used an $N$-body simulation code {\tt GORILLA}
\citep{Tanikawa09}. In this simulation, the first binary is formed
during the interval $18.323t_{\rm rh,i}$ -- $18.442t_{\rm rh,i}$,
where $t_{\rm rh,i}$ is the half-mass relaxation time at the initial
time, and the half-mass relaxation time is defined by
\cite{Spitzer71}. We have defined the first binary as a binary whose
binding energy is more than $10kT$, where $3/2kT$ is the average
kinetic energy of cluster stars at the initial time. Additionally,
this binary survives until it escapes from the cluster.

During the interval when the first binary is formed ($18.323t_{\rm
  rh,i}$ -- $18.442t_{\rm rh,i}$), we take snapshots at every
$0.01t_{\rm cr,c}$, where $t_{\rm cr,c}$ is instantaneous core
crossing time. The instantaneous core crossing time is given by
\begin{equation}
  t_{\rm cr,c}=\frac{r_{\rm c}}{v_{\rm c}},
\end{equation}
where $r_{\rm c}$ is the core radius defined by \cite{Casertano85}
with modifications of \cite{McMillan90}, and $v_{\rm c}$ is the
stellar velocity dispersion in the core. Using these snapshots, we
have analysed orbits of stars involved in the first-binary
formation. Here, we introduce $\tau$, which is time scaled by the
current $t_{\rm cr,c}$, expressed as
\begin{equation}
  \tau = \int \frac{dt}{t_{\rm cr,c}}.
\end{equation}
The time $\tau$ is useful for understanding dynamical processes in the
core. We define $\tau=0$ as the time when the first binary is
formed. In terms of $\tau$, we take the snapshots from $\tau=-67.93$
to $\tau=76.17$. In order to reduce the data sizes of these snapshots,
we include only subsets of stars in these snapshots. Generally we
define these subsets as the stars in the core.  If the number of stars
in the core is less than $40$, however, we include the $40$ nearest
stars around the density centre, which is also defined as in
\cite{Casertano85}, in these snapshots.

\begin{figure*}
 \begin{center}
   \includegraphics[scale=1.2]{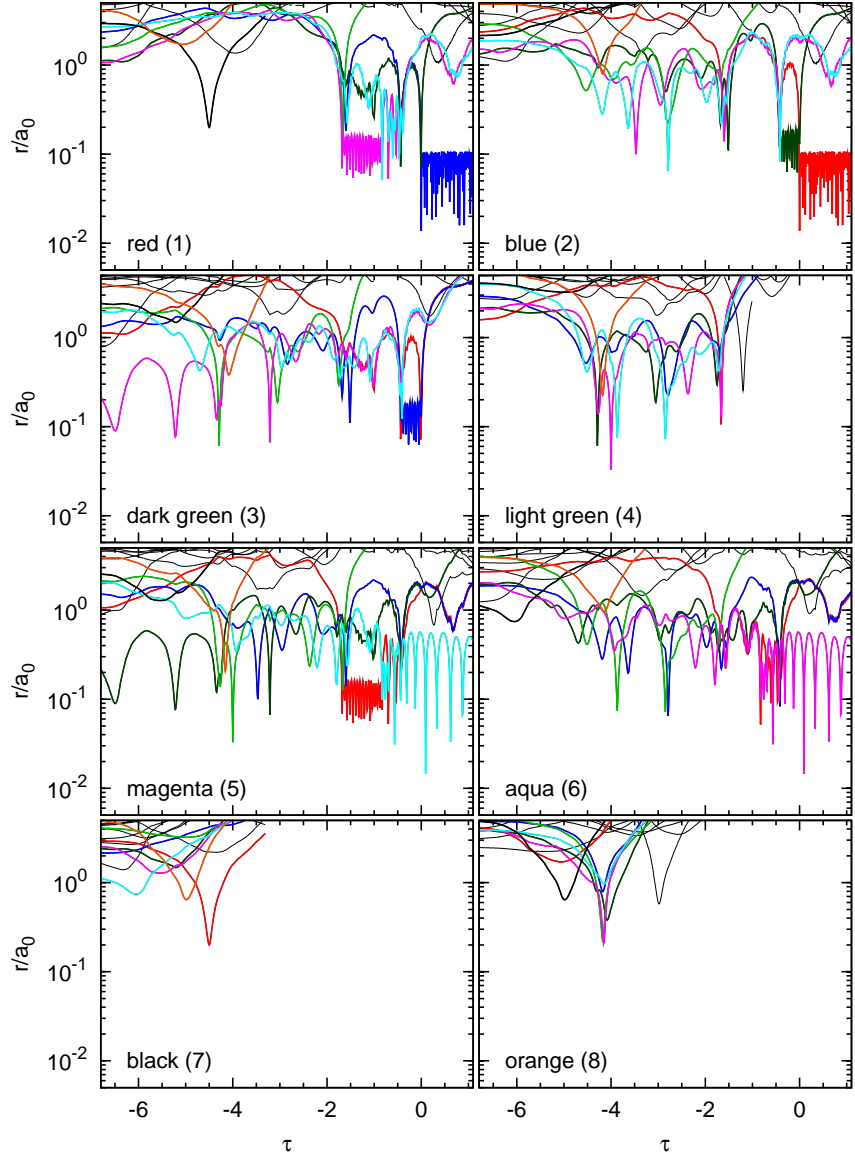}
 \end{center}
 \caption{Time evolution of separations between stars involved in
   binary formation, which is almost the same as fig. 7 in Paper I,
   and is plotted over the same range of scaled time $\tau$.  This
   figure differs from Paper I, however, by the addition of stars 7
   and 8, which are introduced in Sec.\ref{sec:phase_1st}. The
   definitions of $\tau$ and $a_0$ are in the main text.}
 \label{fig:paperI}
\end{figure*}

Figure \ref{fig:paperI} shows the time evolution of separations
between stars involved in binary formation, in units of $a_0$, where
$a_0$ is the semi-major axis of a binary with binding energy
$1kT$. This figure is almost the same as fig. 7 of Paper I. In Paper
I, we drew the orbits of only six stars numbered from 1 to 6.  Here we
have added two more stars, in order to highlight further some of the
earlier stages of these interactions, around time $\tau = -4$.  In
Paper I, the six stars were chosen in the following way.

\begin{figure}
 \begin{center}
   \includegraphics[scale=1.8]{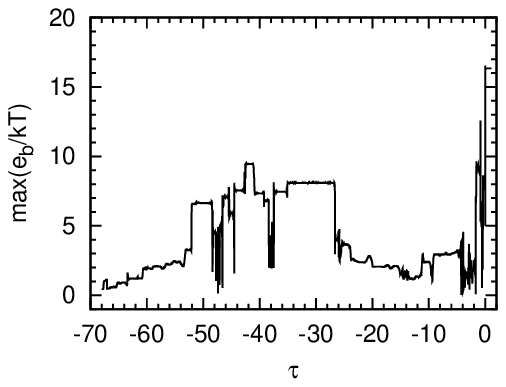}
 \end{center}
 \caption{Time evolution of the maximum binding energy between any two
   stars (among the stars we followed) at each time.}
 \label{fig:ebmax}
\end{figure}

First, we followed the run from the beginning in order to find the
first binary with a total energy of more than $10kT$, which turned out
to be the binary (1,2) at $\tau = 0$. Next, we extended our search
backwards in time, from that point on, to look for stars that had
significant reactions with the final binary components, from $\tau =
0$ back to $-1.6$. We concluded that the hard binary came into being
only at $\tau = -1.6$. In fact, binaries with energy less than $10kT$
are present before $\tau=-1.6$. For example, we can see in figure
\ref{fig:ebmax} that there is a binary with more than $9kT$ at around
$\tau=-38$. However, we do not regard them as the first binary, since
they disappear until $\tau=-4$. Although there are some pairs of stars
with binding energy about $3kT$ during the interval $\tau=-4$ -- $-2$,
they should not be regarded as binaries. This is because their binding
energies are strongly fluctuating. Typical binaries keep their binding
energies constant, such as a binary with binding energy of $\sim 8kT$
during the interval $\tau=-37$ -- $-24.4$.

Second, by considering the distances of other stars from 1 and 2, in
the top row of fig. 7 in Paper I (our figure \ref{fig:paperI}), we
found that stars 3 and 5 also played an important role: the presence
of binaries (1,5) and (2,3) is obvious during the periods $-1.6 < \tau
< -0.8$ and $-0.5 < \tau < 0$, respectively.  In addition, star 4
makes a close encounter to star 1 at $\tau = -1.6$ in the top left
panel (though the details are somewhat crowded) and star 6 can be seen
to dance with stars 1 and 5 in the panels of the third row from $\tau
= -0.8$ onwards.

\begin{landscape}
\begin{table}
\caption{Dynamical interpretation of the first binary formation in
  Paper I and this paper.}
\label{tab:paperI}
\begin{center}
\footnotesize
\begin{tabular}{clcl}
\hline
\multicolumn{2}{c}{Paper I} & \multicolumn{2}{c}{This paper} \\
\hline
\hline
$\tau$ & \multicolumn{1}{c}{Events} & $\tau$ & \multicolumn{1}{c}{Events} \\
\hline
$\lesssim -1.5$ & interaction among stars 1, 2, 4, and 5 leads to &
$-1.6$ & interaction among stars 1, 2, 3, 4, 5, and 6 leads to \\

& formation of binary [1,5] & & formation of binary [1,5] ($9kT$) \\

$\gtrsim -1.0$ & interaction among stars 1, 2, 3, 5, and 6 leads to &
$-0.8$ & invasion of star 6 into binary [1,5] leads to \\

& dissolution of binary [1,5] & & formation of democratic triple
[1,5,6] \\

$ -0.5$ & formation of binary [2,3] from subsystem [1,2,3,5,6] &
$-0.5$ & invasion of stars 2 and 3 into democratic triple [1,5,6]
leads to \\

& & & formation of binaries [2,3] and [5,6] \\

$ 0.0$ & formation of binary [1,2] ($>10 kT$) by direct exchange &
$0.0$ & formation of binary [1,2] ($16 kT$) by direct exchange \\
\hline
\end{tabular}
\end{center}
\end{table}
\end{landscape}

Table \ref{tab:paperI} summarises what can be gleaned from Paper I, in
the left column.  For comparison, we present in the right hand column
somewhat refined information about these events as found in the
current paper. Note that, in the ``exchange'' interactions in
Table~\ref{tab:paperI}, a binary component is replaced by an intruder.

\section{A new analysis of the first run of Paper I}
\label{sec:history}

\subsection{Work Functions}

In Paper I, we showed that more than three stars come close to each
other at the first binary formation. However, we did not quantify how
each star contributes to the binary formation. For this purpose, in
the present paper we use binding energies and ``work functions''. The
former and latter quantify binary formations and contribution of stars
to the binary formations, respectively. Actually, the two quantities
are useful not only for binary formation, but also for binary
evolution, such as hardening, softening, and ionisation. Furthermore,
we can generalise them to subsystems or multiple stars (``tuples'' for
short) which have more than two stars , so that a tuple with two stars
is a binary.

The compactness of a tuple consisting of more than one star may be
quantified by a binding energy. The binding energy of tuple $i$ is
expressed as
\begin{equation}
  E_i(t) = \sum_{k<l}^{n_i}
  \frac{Gm_{i_k}m_{i_l}}{\left|\bm{r}_{i_k}-\bm{r}_{i_l}\right|} -
  \frac{1}{2}\sum_k^{n_i} m_{i_k} \left(\bm{v}_{i_k}^2 - \bm{v}_{\rm
    cm}^2\right), \label{eq:bindingenergy}
\end{equation}
where $n_i$ is the number of components of tuple $i$, and $\bm{v}_{\rm
  cm}$ is the centre-of-mass velocity of tuple $i$, i.e. $\bm{v}_{\rm
  cm}=\left(\sum_k^{n_i} m_{i_k} \bm{v}_{i_k} \right)/
\left(\sum_k^{n_i} m_{i_k} \right)$. If the binding energy $E_i(t)$ is
positive, tuple $i$ is a bound system. We can extend this definition
to a binding energy of a mixture of stars and subsystems. Then, we
replace $m_{i,k}$, ${\bm r}_{i,k}$, and ${\bm v}_{i,k}$ in equation
(\ref{eq:bindingenergy}) by the mass and centre-of-mass position and
velocity of a subsystem, respectively.  In this way we can speak of
the binding energy of a star to a subsystem of other stars, for
example.

The binding energy allows us to identify bound subsystems that are
temporarily almost unperturbed or isolated, as they have roughly
constant binding energy.  This raises the question of how these
periods of roughly constant binding energy begin and end.  Clearly, a
significant amount of energy exchange between the subsystem and its
surroundings is involved, at the beginning and the end of each such
period.  To characterise the energy exchange, we can look at the
amount of work done during each event.

We start by determining the rate of energy exchange, defined as the
amount of energy per unit time that the tuple $i$ receives from a
given star $j$:
\begin{equation}
  \dot{E}_{i,j}(t) = -\sum_k^{n_i}\left[\bm{f}_{i_k,j} \cdot
    \left(\bm{v}_{i_k} - \bm{v}_{\rm cm}\right)\right],
  \label{eq:workfunction}
\end{equation}
where $\bm{f}_{i_k,j}$ is the force exerted by star $j$ on component
$k$ of tuple $i$; this can be expressed as
$\bm{f}_{i_k,j}=-Gm_jm_{i_k}
(\bm{r}_{i_k}-\bm{r}_j)/|\bm{r}_{i_k}-\bm{r}_{j}|^3$.  We call this
quantity a ``power function''.  Note, incidentally, that the
``component'' may itself be a subsystem, as before.

In order to find out which star or stars are responsible for a
transition event we can compute the power function for various
candidate stars $j$.  But since the event is identified by means of
the change in binding energy of tuple $i$, it is easier and more
reliable to compare the {\sl integral} of the power function with the
binding energy of tuple $i$.  We refer to this as a ``work function''.
In practice we plot the graph of $E_i(t)$ along with the graphs of
$E_{i,j}(t)$ for several candidate stars $j$.  An example is the lower
panel of figure \ref{fig:work_0th_1}, which will be discussed in
Section \ref{sec:phase_1st}.

\subsection{Overview of the history for the binary formation}
\label{sec:overview}

\begin{figure*}
 \begin{center}
   \includegraphics[scale=0.75]{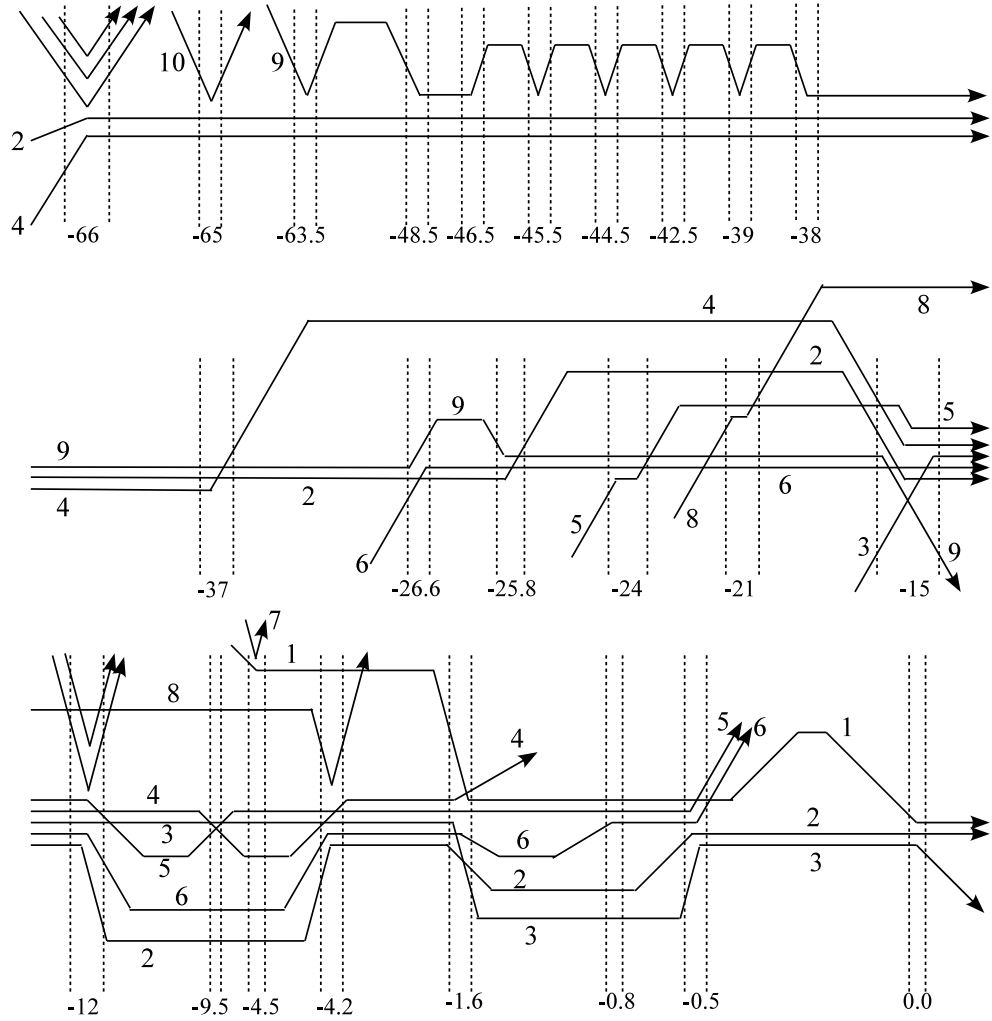}
 \end{center}
 \caption{Illustration of interactions involved in the first binary
   formation. Close parallel lines indicate (schematically) temporary
   bound subsystems.}
 \label{fig:diagram}
\end{figure*}

As we show in the remainder of this section, these diagnostics can be
used to unravel the main interactions which eventually give rise to
the first $10kT$ binary.  On this basis it is possible to construct
(manually) a schematic but detailed graphic description of these
interactions (figure \ref{fig:diagram}), in analogy with a diagram for
a resonant three-body interaction presented by
\citet[fig.3]{1983ApJ...268..319H}.  We sometimes refer to this as a
kind of Feynman diagram, in analogy with somewhat similar figures used
in perturbative quantum field theories.  Though it depends on results
which are still to be presented, it will aid the reader to follow the
analysis with reference to this diagram.

In this reanalysis we divide the evolution into four phases: $-6.75 <
\tau < -4.2$, $-4.2 < \tau < -1.6$, $-1.6 < \tau < 0.0$, and $0.0 <
\tau < 1.1$.  The start and end points merely delimit the range of
times which were considered in detail in Paper I (see fig. 7 in Paper
I, and figure \ref{fig:paperI} in the present paper). The other times
are significant events identified in section \ref{sec:summaryI} and
table \ref{tab:paperI}. These four phases are considered in the
following sub-sections, and the phase before $\tau = -6.75$ is
analysed in section \ref{sec:prehistory}.

\subsection{The Era $-6.75 < \tau < -4.2$}
\label{sec:phase_1st}

\begin{figure}
 \begin{center}
   \includegraphics[scale=1.8]{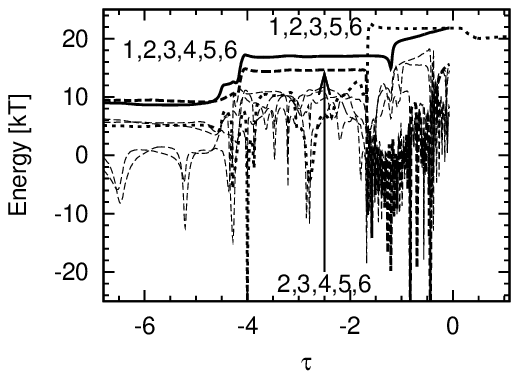}
 \end{center}
 \caption{Time evolution of binding energies of the sextet
   (1,2,3,4,5,6) and any tuple of quintets in the sextet. Thick solid,
   dashed, and dotted curves indicate the binding energies of the
   sextet, the quintet (2,3,4,5,6), and the quintet (1,2,3,5,6),
   respectively, and thin dashed curves those of the other quintets.}
 \label{fig:tuple_0th}
\end{figure}

\begin{figure}
 \begin{center}
   \includegraphics[scale=1.8]{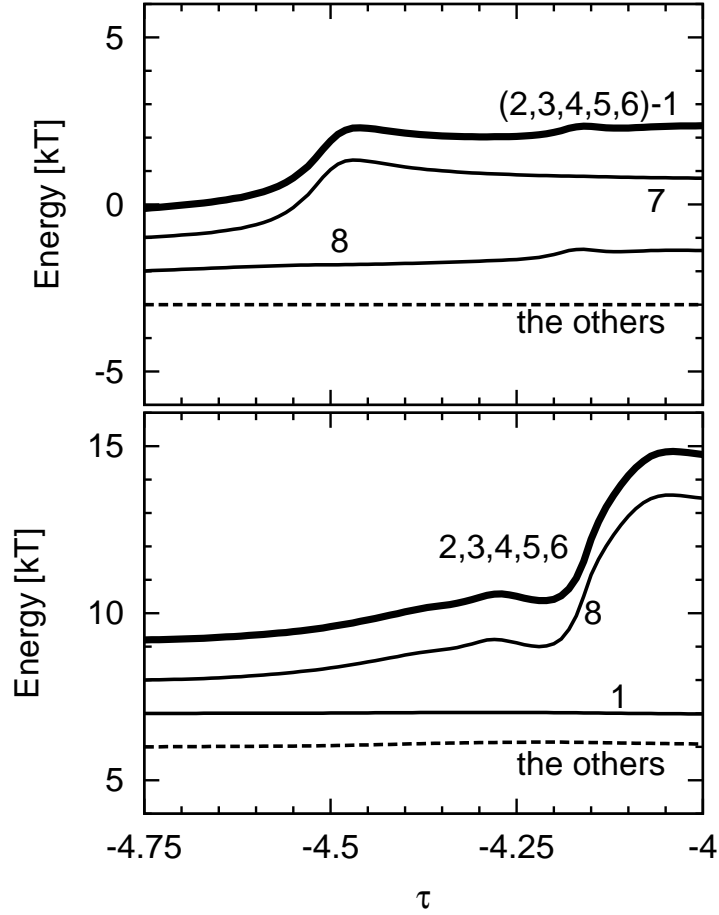}
 \end{center}
 \caption{Time evolution of binding energies between a quintuple
   system (2,3,4,5,6) and star 1 (upper panel), and of the quintuple
   system (2,3,4,5,6) (lower panel), and of work done on them. The
   binding energies and work done are indicated by thick and thin
   curves, respectively. Numbers beside curves for the work indicate
   which stars do the work. The work is integrated from the time $\tau
   = -4.75$. In order to be eye-friendly, the initial values of the
   work done are not zero, which applies also for figures
   \ref{fig:work_2nd}, \ref{fig:work_3rd}, \ref{fig:work_5th},
   \ref{fig:m1st_p1}, and \ref{fig:m1st_p3}.}
 \label{fig:work_0th_1}
\end{figure}

Turning to the analysis of the first phase, $-6.75 < \tau < -4.2$, we
see from the solid curve in figure \ref{fig:tuple_0th} that system
(1,2,3,4,5,6) is not isolated, as its binding energy changes abruptly
at about $\tau = -4.5$ and $\tau = -4.2$. This is the reason why we
include two additional stars in figure \ref{fig:paperI}. At the former
time, figure \ref{fig:paperI} (or fig. 7 in Paper I) shows a close
approach to star 1 by a star which was unidentified in Paper I, but
which we now label as star 7.  In this encounter, the top panel of
figure \ref{fig:work_0th_1} shows that work is done by star 7 on a
binary consisting of star 1 and the quintuple system (2,3,4,5,6).  In
this encounter, star 1 becomes bound to the quintuple system
(2,3,4,5,6).

\begin{figure}
 \begin{center}
   \includegraphics[scale=1.8]{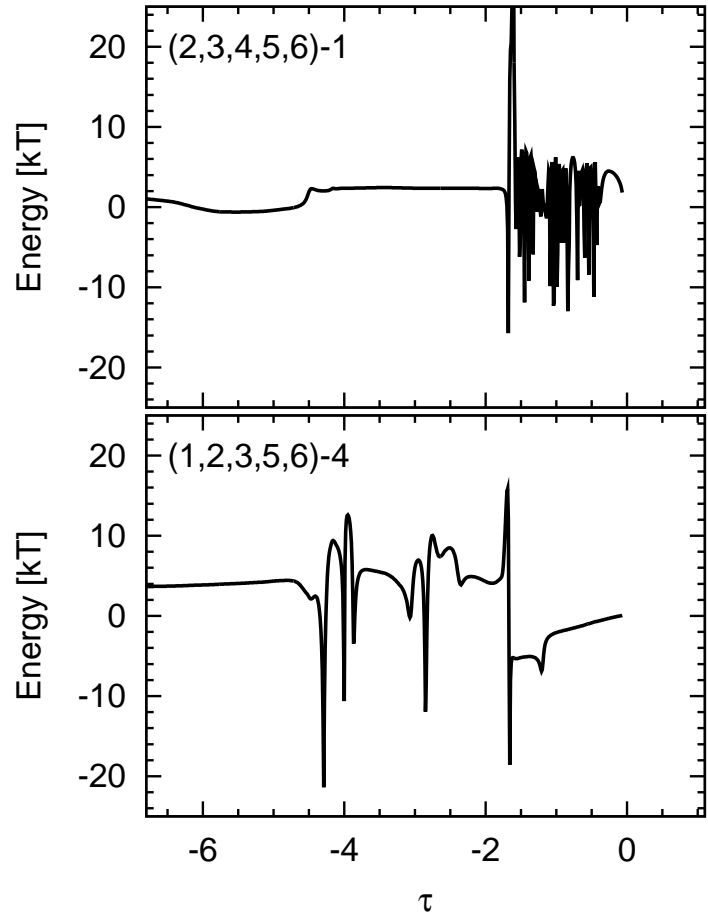}
 \end{center}
 \caption{Time evolution of the binding energies between the quintet
   (2,3,4,5,6) and star 1 (top), and between the quintet (1,2,3,5,6)
   and star 4 (bottom).}
 \label{fig:twog_0th}
\end{figure}

\begin{figure}
 \begin{center}
   \includegraphics[scale=1.55]{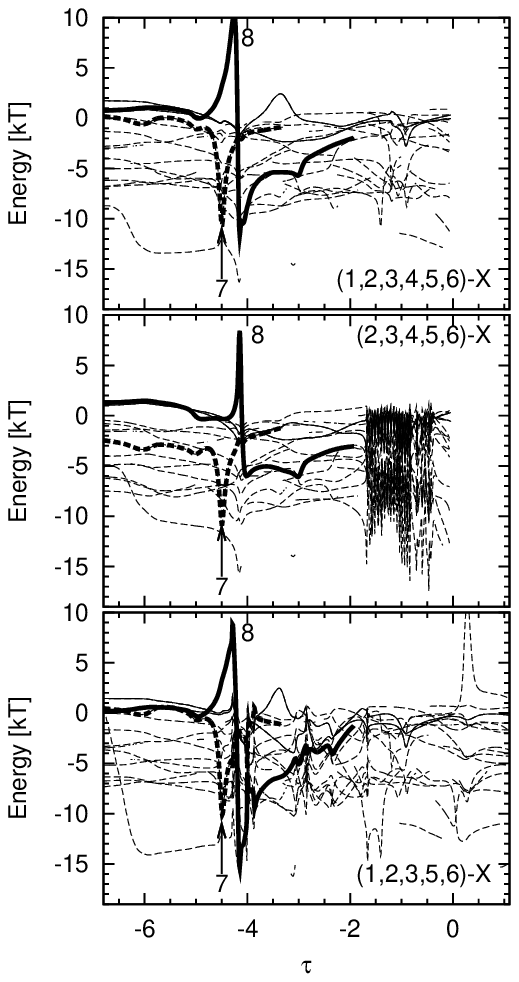}
 \end{center}
 \caption{Time evolution of binding energies between a star ``X'' and
   either the sextet (1,2,3,4,5,6) (top) or one of the quintets
   (2,3,4,5,6) (middle) and (1,2,3,5,6) (bottom). The star ``X'' is
   usually unnumbered.  Two stars which are particularly bound to the
   sextet (or quintet (2,3,4,5,6)) are indicated by solid curves: the
 star 8 is shown by a thick solid curve, and star 7 by a thick dashed
 curve; the other stars are shown by thin dashed curves. The distances
 of the stars 7 and 8 are shown in figure \ref{fig:paperI}.}
 \label{fig:seventhman}
\end{figure}

In the phase up to this event at $\tau = -4.5$, five stars of the
sextet components (stars 2, 3, 4, 5, and 6) are bound with a binding
energy of about $10kT$ (see the thick dashed curve in figure
\ref{fig:tuple_0th}), and the five stars compose a quintuple system
(2,3,4,5,6).  Star 1 is unbound to the quintuple system, as can be
seen in the top panel of figure \ref{fig:twog_0th}.  The quintuple
system is unperturbed until the event at $\tau = -4.5$, since its
binding energy is kept constant. Actually, three hitherto unnumbered
stars are weakly bound to the quintuple system at $\tau = -6.75$, but
they have gradually become unbound  by $\tau = -5$ (see the middle
panel of figure \ref{fig:seventhman}). They are not members of the
quintuple system.  We conclude that the quintuple system (2,3,4,5,6)
is nearly isolated from the other stars during this phase up to
$\tau=-4.5$, while star 1 is unbound to the quintuple system.

It is important to confirm that no other stars are involved in the
binding of the system (2,3,4,5,6),1 at $\tau=-4.5$, and the hardening
of the quintuple system (2,3,4,5,6) at $\tau=-4.2$. In both cases the
work function by other stars included in snapshots, which we
take in a way described in section~\ref{sec:summaryI}, varies
only a little (see dashed curves in both of the panels of figure
\ref{fig:work_0th_1}).

\begin{figure}
 \begin{center}
   \includegraphics[scale=1.5]{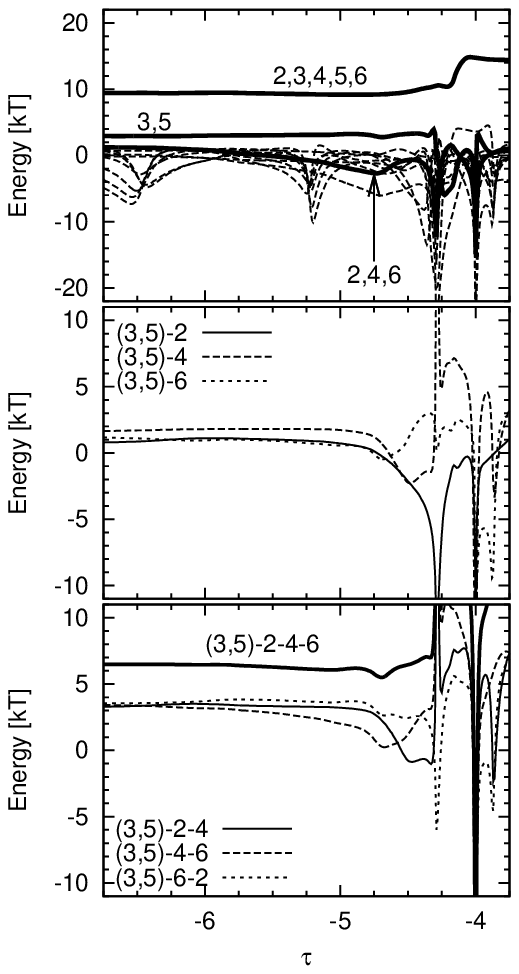}
 \end{center}
 \caption{(Top) Time evolution of binding energies among a quintet
   (2,3,4,5,6) and a trio (2,4,6), and between any pair in the
   quintet. Thick curves show the binding energies among the quintet
   and the trio, and between a pair (3,5), and dashed curves those
   between all the pairs, except the pair (3,5). (Middle) Time
   evolution of binding energies between the pair (3,5) and either one
   of stars 2, 4, and 6. (Bottom) Time evolution of binding energies
   among the binary (3,5) and all of stars 2, 4, and 6, and among the
   binary (3,5) and any two of the stars 2, 4, and 6.}
 \label{fig:tuple_1st}
\end{figure}

Now we investigate the substructures of the quintuple system
(2,3,4,5,6) described above.  First, as seen in the top panel of
figure \ref{fig:tuple_1st}, there is a binary (3,5) with about $3kT$,
which corresponds to a pair of stars with binding energy $3kT$ during
$\tau=-9.5$ -- $-4.2$, seen in figure \ref{fig:ebmax}. The other three
stars compose no binary. These three stars are not bound to each other
as a democratic triple system (see the lowest solid black curve in the
top panel of figure \ref{fig:tuple_1st}). Thus the quintuple system
(2,3,4,5,6) consists of four components: one binary (3,5), and three
single stars 2, 4, and 6.  These four components are bound with energy
$6kT$ (see the bottom panel of figure \ref{fig:tuple_1st}).

Next, we analyse how these four components are structured. We can see
from the middle panel of figure \ref{fig:tuple_1st} that pairs between
the binary (3,5) and either of the other three stars have positive
binding energies, and that their binding energies are kept almost
constant.  This means that the binary (3,5) and the other three stars
compose hierarchical triple systems individually, and that they are
not perturbed much by each other.

The binary (3,5) plays an important role in binding the four
components which are obtained by choosing any three of the four
objects (3,5), 2, 4 and 6. Three of these four components have
positive binding energies, i.e.  the three which include the binary
(3,5) (see the bottom panel of figure \ref{fig:tuple_1st}). On the
other hand, the three single stars are unbound (see the top panel of
figure \ref{fig:tuple_1st}). Therefore, the configuration of these
four components is not democratic, but similar to a planetary system;
the binary and single stars correspond to a sun and planets,
respectively.

\begin{figure}
 \begin{center}
   \includegraphics[scale=1.8]{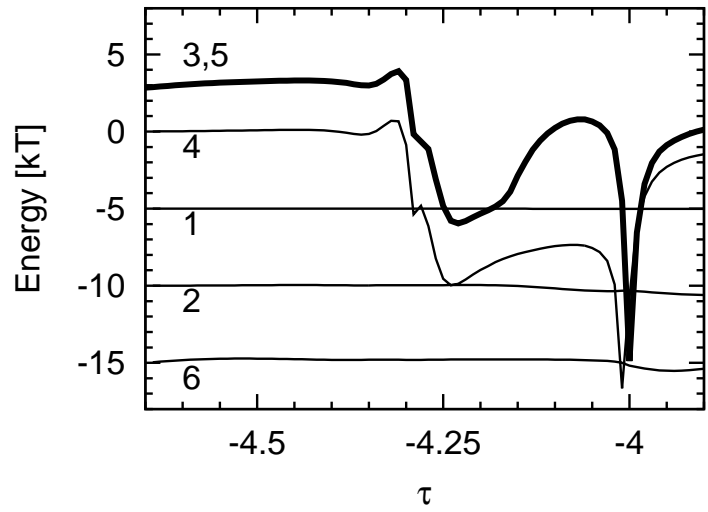}
 \end{center}
 \caption{Time evolution of the binding energy of the binary (3,5)
   (thick curve), and work done by the four numbered stars 1, 2, 4,
   and 6 on this binary (thin curves). Numbers beside the work curves
   indicate the relevant star. The work is integrated from the time
   $\tau = -4.65$.}
 \label{fig:work_2nd}
\end{figure}

In summary, the quintuple system (2,3,4,5,6) has four components: a
binary (3,5) and three single stars, in a configuration analogous to a
planetary system. This configuration breaks down soon, however, since
the mass ratios of the binary to the single stars are not large.  In
fact the binary (3,5) is ionised at the end of the current phase.
This binary is destroyed by star 4. As seen in figure
\ref{fig:work_2nd}, star 4 does work on the binary (3,5) at the moment
when its binding energy becomes negative. No other stars do work on
the binary (3,5).

Now we describe the other events around the close of the phase,
i.e. $\tau = -4.2$.  At this time the binding energies among the
sextuple component (1,2,3,4,5,6) and among the quintet component
(2,3,4,5,6) become larger (see figure \ref{fig:tuple_0th}), while the
binding energy between the quintuple system (2,3,4,5,6) and star 1 is
kept almost constant (see the top panel of figure
\ref{fig:twog_0th}). This means that the quintuple system (2,3,4,5,6)
becomes bound more tightly. This event results from work done by a
hitherto unnamed star which we subsequently refer to as star 8, and
which intrudes into the quintuple system (2,3,4,5,6) at the time of
the event (see figure \ref{fig:paperI}).  From the bottom panel of
figure \ref{fig:work_0th_1}, we see that star 8 is almost entirely
responsible for the increase in binding energy of the quintuple around
$\tau= - 4.2$.  At the same time star 8 slightly hardens the binary
consisting of star 1 and the quintuple system (2,3,4,5,6) (see the top
panel of figure \ref{fig:work_0th_1}). However, its effect is small,
compared to that of star 7 a short time earlier.

\subsection{The Era $-4.2 < \tau < -1.6$}
\label{sec:phase_2nd}

As we have seen, by the start of the second phase, star 1 has joined
the quintuple system (2,3,4,5,6).  Throughout the second phase the
binding energy between star 1 and the quintuple system is positive,
about $2kT$, and constant (see the top panel of figure
\ref{fig:twog_0th}).  The sextet (1,2,3,4,5,6) forms a bound sextuple
system, and is unperturbed by other stars (figure
\ref{fig:tuple_0th}).  Indeed no other star is continuously bound to
this sextuple system (see the top panel of figure
\ref{fig:seventhman}).  Since the binding energy of the quintuple
system (2,3,4,5,6) is also constant (see figure \ref{fig:tuple_0th}),
the quintuple system is unperturbed by other stars including star 1,
as star 1 is far from the other numbered stars (see figure
\ref{fig:paperI}).  In practice, the quintuple system (2,3,4,5,6)
survives undisturbed from the previous phase.

\begin{figure}
 \begin{center}
   \includegraphics[scale=1.6]{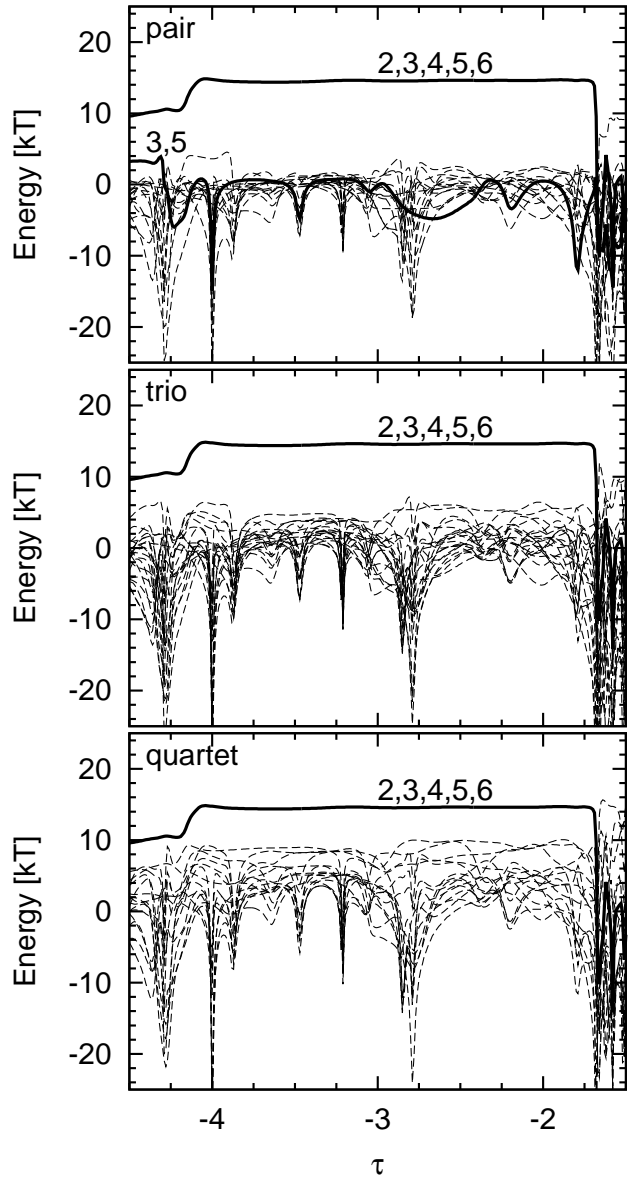}
 \end{center}
 \caption{Time evolution of binding energies of a quintet (2,3,4,5,6)
   (each panel) and any pair (top), any trio (middle), and any quartet
   (bottom) in the quintet. All these pairs, trios, and quartets are
   indicated by dashed curves, except the pair (3,5) (solid curve).}
 \label{fig:tuple_2nd}
\end{figure}

In order to investigate the internal structure of the quintuple
system, we focus on all the tuples consisting of its components.
After the soft binary (3,5) is destroyed at $\tau \sim - 4.25$, no
persistent substructure is formed until the end of this phase, at
$\tau = -1.6$. As seen in all the panels of figure
\ref{fig:tuple_2nd}, no binding energy of any tuple keeps constant
during this phase. Occasionally, the binding energies of some tuples
are temporarily positive. However, their lifetime is less than unity
in the units of $\tau$, which is similar to the crossing time of the
quintuple system (2,3,4,5,6). We conclude that the quintuple system
has a democratic configuration, and therefore we can write the
configuration of the sextuple system as [(2,3,4,5,6),1]. This is
consistent with the conclusion, described in section
\ref{sec:summaryI}, that the hard binary came into being only at
$\tau=-1.6$.

At $\tau \sim -1.6$, the end of the second phase, or the beginning of
the third phase, the sextuple system (1,2,3,4,5,6) changes
dramatically.  Star 1 intrudes into the quintuple system (2,3,4,5,6),
an interaction which results in the ejection of star 4 from the
sextuple system (1,2,3,4,5,6). During this interaction, the sextet
(1,2,3,4,5,6), which is bound before $\tau = -1.6$, is unperturbed by
the other cluster stars, since the binding energy among the sextet
components is constant before and (for a short time) after this
interaction (see figure \ref{fig:tuple_0th}).  On the other hand, star
4 becomes unbound to the new quintuple system (1,2,3,5,6): the binding
energy between star 4 and the new quintuple system becomes and remains
essentially negative (see the bottom panel of figure
\ref{fig:twog_0th}).  From the right second-row panel of figure
\ref{fig:paperI}, we can also see that star 4 recedes from all the
components of the new quintuple system.

\subsection{The Era $-1.6 < \tau < -0.8$}
\label{sec:phase_3rd}

In this section, we focus on the quintuple system (1,2,3,5,6) and its
internal structure. In the third phase $-1.6<\tau<-0.8$, and indeed
throughout the interval $-1.6 < \tau < 0.0$, the new quintet members
(1,2,3,5,6) are bound to each other. In fact the binding energy of the
quintuple system is more than $20kT$, as can be seen in figure
\ref{fig:tuple_0th}. It is almost isolated, since its binding energy
keeps constant during this phase.  No other star is continuously bound
to this quintuple system (see the bottom panel of figure
\ref{fig:seventhman}). Star 4 was ejected from the sextet
(1,2,3,4,5,6) in the creation of the new quintet, and it can be seen
in the lower panel of figure \ref{fig:twog_0th} that the binding
energy between star 4 and the quintuple system is perturbed by an
event at $\tau = - 1.25$.  This is caused by an encounter with an
unnamed star (figure \ref{fig:paperI}), which also affects the binding
energy of the sextet (figure \ref{fig:tuple_0th}).

\begin{figure}
 \begin{center}
   \includegraphics[scale=1.5]{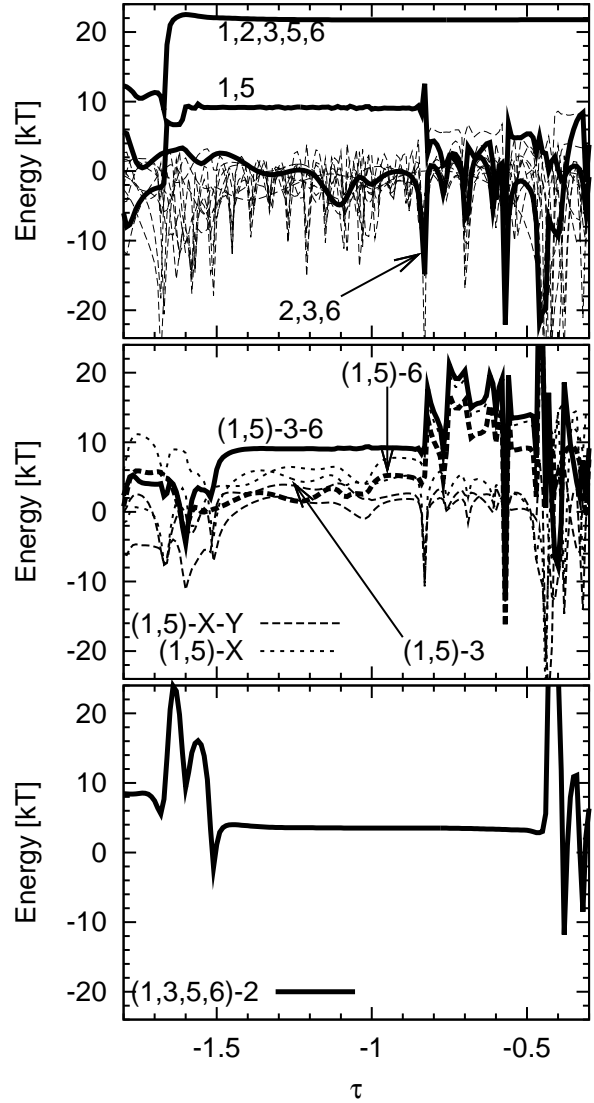}
 \end{center}
 \caption{(Top) Time evolution of binding energies between any pair of
   components in the quintet (1,2,3,5,6), and among the trio
   (2,3,6). Dashed curves indicate binding energies between all the
   pairs, except the pair (1,5). (Middle) Time evolution of binding
   energies between the pair (1,5) and any one of the stars 2, 3, and
   6, and among the pair (1,5) and any two of the stars 2, 3, and
   6. (Bottom) Time evolution of binding energies between the quartet
   (1,3,5,6) and star 2.}
 \label{fig:tuple_3rd}
\end{figure}

First, we search for substructures which consist of binary
stars. There is only one, a binary (1,5) with energy $9kT$ (see the
top panel of figure \ref{fig:tuple_3rd}), and the other three stars do
not compose any binary or triple system (see the top panel of figure
\ref{fig:tuple_3rd}).  Next, we seek stable substructures which
contain the binary (1,5) and single stars 2, 3, and 6. We can see from
the middle panel of figure \ref{fig:tuple_3rd} that the binding energy
among the binary (1,5) and two single stars 3 and 6 remains constant
during $-1.5<\tau<-0.8$. This means that the quadruple system
(1,3,5,6) is unperturbed by star 2. This quadruple system is bound to
star 2, and the pair consisting of this quadruple system and star 2 is
unperturbed by the other stars (see the bottom panel of figure
\ref{fig:tuple_3rd}).

Now we focus attention on the internal structure of the quadruple
system (1,3,5,6) containing the binary (1,5) and two single stars 3
and 6. The binding energy of pair (3,6) is
negative (see the top panel of figure \ref{fig:tuple_3rd}), and those
between the binary (1,5) and either one of stars 3 and 6 are positive
(see the middle panel of figure \ref{fig:tuple_3rd}). This means that
they compose a planetary system ; the binary (1,5) is a sun and stars
3 and 6 are planets, similarly to the quintuple system (2,3,4,5,6)
during $\tau<-4.2$. However, the binding energies between the binary
(1,5) and either one of stars 3 and 6 are fluctuating, and therefore
the ``planets'' perturb each other.

In summary, the structure of the quintuple system (1,2,3,5,6) may be
summarised as \{[(1,5),3,6],2\}. Note that the quadruple system
(1,3,5,6) is structured as a sort of planetary systems in which the
binary (1,5) is a sun, and stars 3 and 6 are planets; they are {\it
  not}, however, a democratic triple system with respect to the binary
(1,5) and single stars 3 and 6.

\begin{figure}
 \begin{center}
   \includegraphics[scale=1.8]{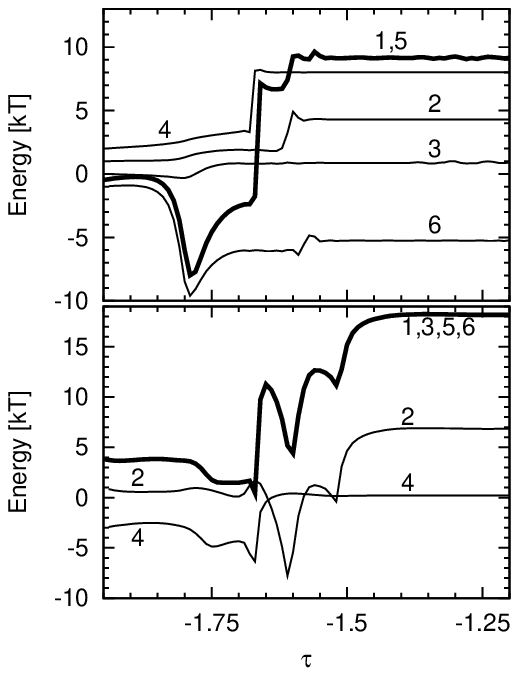}
 \end{center}
 \caption{Time evolution of the binding energies of a binary (1,5) and
   a quadruple system (1,3,5,6), and of work done on these
   substructures by the corresponding numbered star.  The work is
   integrated from the time $\tau = -1.95$.}
 \label{fig:work_3rd}
\end{figure}

We now consider the formation of the new substructures which
originated near the start of this phase, i.e. the binary (1,5) and the
quadruple system (1,3,5,6). The top panel of figure \ref{fig:work_3rd}
shows the evolution of the binding energy of the binary (1,5) and the
work done by stars 2, 3, 4, and 6 on this binary. We do not need to
consider the work done by the other stars; the sextuple system
(1,2,3,4,5,6) is isolated during this phase, which can be seen in the
constancy of the binding energy of the sextuple system (1,2,3,4,5,6)
throughout the current phase (see figure \ref{fig:tuple_0th}).

As seen in the top panel of figure \ref{fig:work_3rd}, the binding
energy of the binary (1,5) is mainly increased by work done by star
4. Stars 2 and 6 also contribute to the increase of its binding energy
after the binary (1,5) has become hard ($\sim 8kT$). However, they
perturb the binary (1,5) only marginally.  We also see the binding
energy of the quadruple system (1,3,5,6) plotted in the bottom panel
of figure \ref{fig:work_3rd}. We can say that the quadruple system is
formed and hardened by work done by star 4 at $\tau=-1.65$, and by
work done by star 2 at $\tau=-1.5$, two events which help to isolate
the quadruple from external disturbance.

\subsection{The Era $-0.8 < \tau < -0.5$}
\label{sec:phase_4th}

In this phase, the quintuple system (1,2,3,5,6) survives without
external disturbance from the previous phase (see the top panel of
figure \ref{fig:tuple_3rd}).  However, its internal structure is
changed.

\begin{figure}
 \begin{center}
   \includegraphics[scale=1.8]{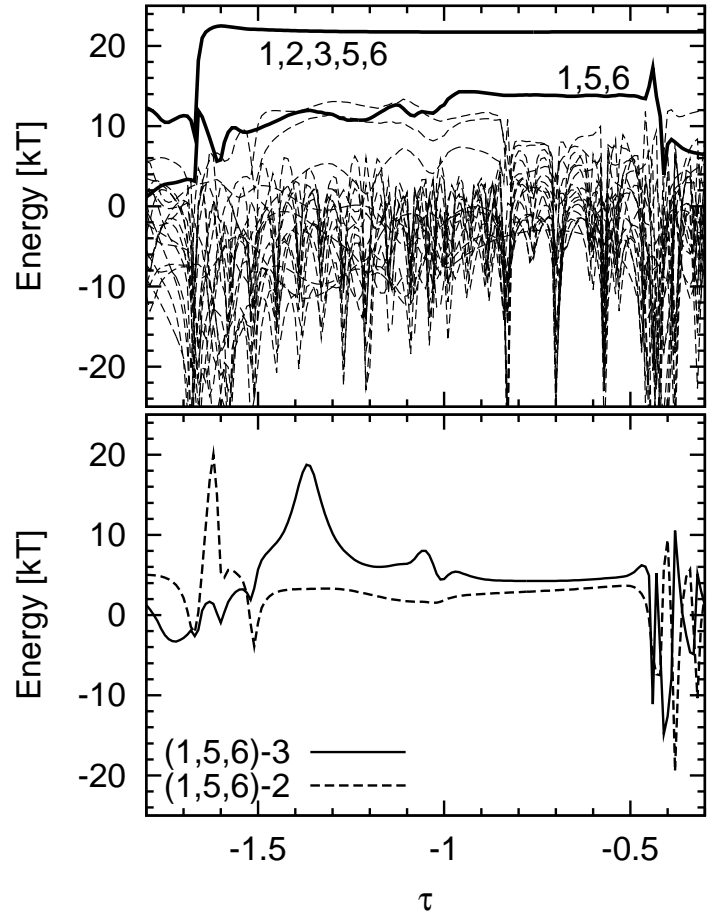}
 \end{center}
 \caption{(Top) Time evolution of binding energies of a quintet
   (1,2,3,5,6) (a solid curve) and any trio in this quintet. The trio
   (1,5,6) is indicated by a solid curve, and the other trios by
   dashed curves. (Bottom) Time evolution of binding energies between
   the trio (1,5,6) and either one of stars 2 and 3.}
 \label{fig:tuple_4th}
\end{figure}

First we consider substructures consisting only of two or three
stars. There is no persistent binary in the quintuple system (see the
top panel of figure \ref{fig:tuple_3rd}), but there is a triple system
(1,5,6) (see the top panel of figure \ref{fig:tuple_4th}). This triple
system is democratic, since no pair in this triple system is bound by
themselves (see the top panel of figure \ref{fig:tuple_3rd} again).

The democratic triple system (1,5,6) is bound to both stars 2 and 3.
The pair between the triple system and star 3 is unperturbed by the
other stars (see the bottom panel of figure \ref{fig:tuple_4th}). On
the other hand, the pair between the triple system and star 2 is
slightly perturbed: note that the binding energy of the pair continues
to increase slowly throughout this phase (see the bottom panel of
figure \ref{fig:tuple_4th}). This pair may be perturbed by star 3,
since the separation between the triple system and star 2 is larger
than that between the triple system and star 3; this can be seen with
a little difficulty in figure \ref{fig:paperI}, and also from the fact
that star 3 is more tightly bound to the triple than star 2 is (the
lower panel of figure \ref{fig:tuple_4th}).  The assumption that no
external star is involved is supported by the bottom panel of figure
\ref{fig:tuple_3rd}, which shows that the quadruple (1,3,5,6) and star
2 are bound to each other and unperturbed. In summary, this discussion
shows that the structure of the quintuple system may be written as
\{[(1,5,6),3],2\}.

The start of the present phase is marked by the disruption of binary
(1,5), as can be seen in the top panel of figure \ref{fig:tuple_3rd}.
It is clear that this event is caused by the invasion of star 6 into
the binary (1,5) , and results in the formation of the democratic
triple system (1,5,6). We do not show the work function of star 6 for
the binary (1,5), but the close interaction of these three stars at
$\tau = -0.8$ is obvious enough in figure \ref{fig:paperI}.  From the
point of view of our theoretical understanding, this is one of the
most significant events in the entire evolution (see section
\ref{sec:dem-res}).

\subsection{The Era $-0.5 < \tau < 0.0$}
\label{sec:phase_5th}

\begin{figure}
 \begin{center}
   \includegraphics[scale=1.45]{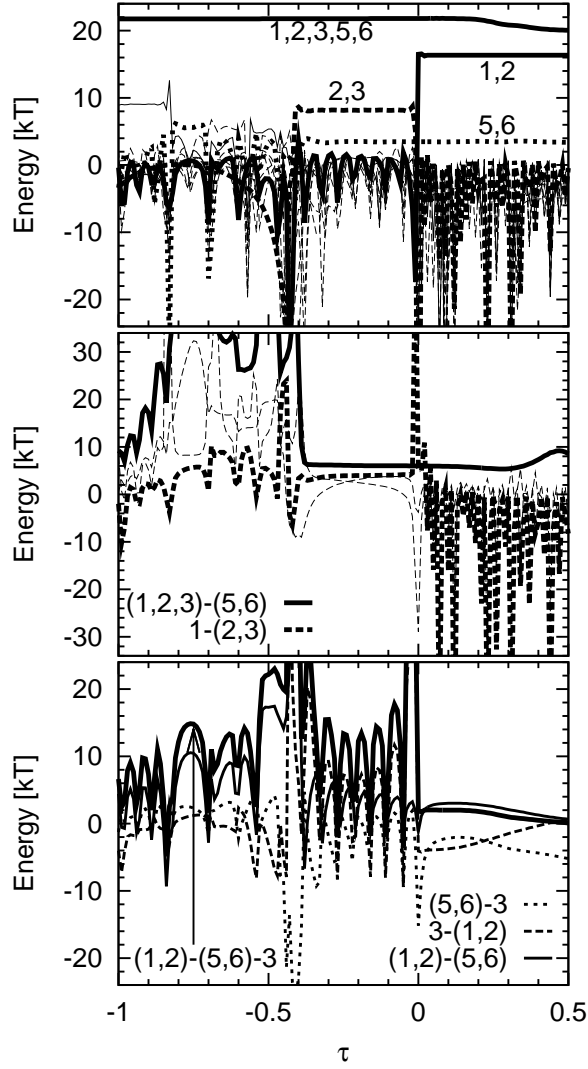}
 \end{center}
 \caption{Time evolution of the quintuple system (1,2,3,5,6) and its
   substructure. (Top) Binding energy between any pair is shown: pairs
   (1,2) (thick solid curve), (2,3) (thick dashed curve), (5,6) (thick
   dotted curve), and the others (thin dashed curves). (Middle)
   Binding energies between any pair of three components: pairs (2,3)
   and (5,6), and star 1, and that between the trio (1,2,3) and the
   pair (5,6). The pairs of the two binaries (2,3) and (5,6) and of
   the binary (5,6) and star 1 are indicated by thin dashed curves.
   (Bottom) Binding energy among three components: pairs (1,2) and
   (5,6) and star 3, and those between any pair of the three
   components.}
 \label{fig:tuple_5th6th}
\end{figure}

In this phase, which strictly begins nearer $\tau = -0.4$, the
quintuple system (1,2,3,5,6) persists from the previous phase, without
perturbation by the other stars (see its binding energy in the top
panel of figure \ref{fig:tuple_5th6th}). However, the internal
structure is greatly changed. As seen in the top panel of figure
\ref{fig:tuple_5th6th}, a hard binary (2,3) and soft binary (5,6) are
formed. Their binding energies are, respectively, $8kT$ and
$3kT$. Therefore, the five stars compose three components containing
less than three stars each: the hard binary (2,3), soft binary (5,6),
and a single star 1.

We now investigate binding energies between each pair of the above
three components (see the middle panel of figure
\ref{fig:tuple_5th6th}). We observe first that the binding energy
between the binary (2,3) and star 1 is almost constant; in other
words, they have a hierarchical configuration. The hierarchical triple
system [(2,3),1] and the soft binary (5,6) are bound to each other
(see also the middle panel of figure \ref{fig:tuple_5th6th}). We can
abbreviate the structure of the quintuple system (1,2,3,5,6) as
\{[(2,3),1],(5,6)\}.

Now we describe how the structure of the previous phase is transformed
into the new structure, i.e. the transformation from \{[(1,5,6),3],2\}
into \{[(2,3),1],(5,6)\}.  We have to consider in particular the
destruction and formation of the innermost structures: the destruction
of the democratic triple system (1,5,6), and the formation of the hard
binary (2,3) and the soft binary (5,6).

\begin{figure}
 \begin{center}
   \includegraphics[scale=1.6]{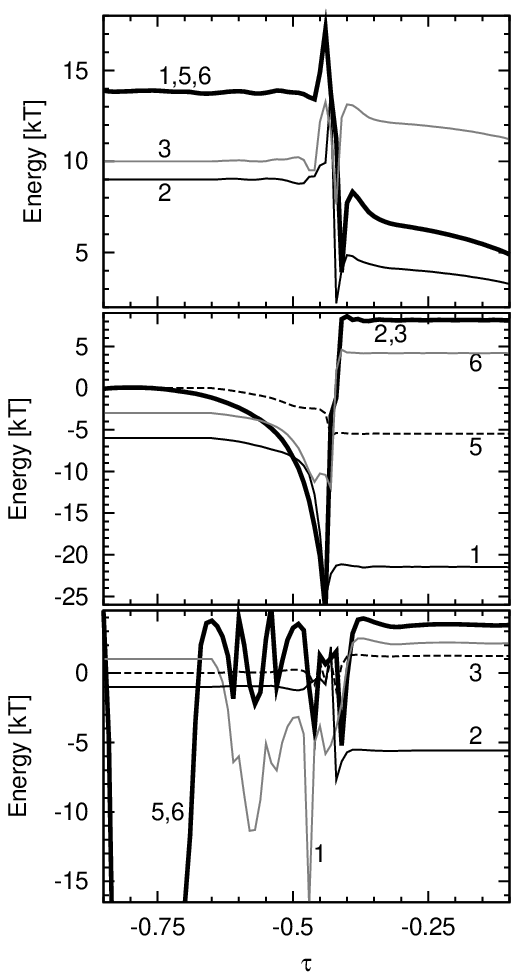}
 \end{center}
 \caption{Time evolution of the binding energies of the trio (1,5,6)
   and pairs (2,3) and (5,6), and of work done on these substructures
   by the star identified by the label numbers.  The work is
   integrated from the time $\tau = -0.65$.}
 \label{fig:work_5th}
\end{figure}

The democratic triple system is destroyed equally by both stars 2 and
3, which were its companions in the previous phase (see the top panel
of figure \ref{fig:work_5th}). The triple also does work on stars 2
and 3, causing them to form a hard binary. In fact the work which
forms this binary is mainly done by star 6, and marginally by star 1
(see the middle panel of figure \ref{fig:work_5th}).

Finally, we investigate the formation of the soft binary (5,6). As
seen in the bottom panel of figure \ref{fig:work_5th}, the work of
stars 1, 2 and 3 on the soft binary (5,6) is complicated.  However, it
is clear that the work of star 1 contributes strongly to the binding
of the soft binary (5,6).  After the work done by star 1, the binding
energy of the two stars 5 and 6 becomes positive. Note that a binary
becomes more bound, i.e. has larger binding energy, if a given star
works on the binary. The definitions of a binary's binding energy and
work function can be seen in equations~(\ref{eq:bindingenergy}) and
(\ref{eq:workfunction}), respectively.

\subsection{The Era $\tau > 0.0$}
\label{sec:phase_6th}

In the final phase, the structure becomes simple.  As can be seen in
the top panel of figure \ref{fig:tuple_5th6th}, there is one hard
binary (1,2) with binding energy $16kT$, and one soft binary
(5,6). The formation of the harder binary was the event which
determined the origin of scaled time $\tau$ in Paper I.  The soft
binary lives from the previous phase. Thus the quintet (1,2,3,5,6) has
three components: two binaries (1,2) and (5,6) and one single star
3. These three components eventually become unbound from each other.
As can be seen in the bottom panel of figure \ref{fig:tuple_5th6th},
the binding energies between star 3 and either one of the binaries
(1,2) and (5,6) are negative. The two binaries are bound at the
beginning of this phase, but their binding energy reaches zero at time
$\tau = 0.5$. The binding energy among the three components behaves in
the same way as that between the two binaries (1,2) and (5,6).
Indeed, in the phase $\tau > 0$, the binding energy of the quintuple
system (1,2,3,5,6) itself begins to change (see the top panel of
figure \ref{fig:tuple_5th6th}).  As we shall see, the quintuple system
is destroyed due to interactions among its components themselves at
$\tau = 0$, and gradually the quintuple system becomes more easily
perturbed by other stars.

In summary, the hard binary (1,2), the soft binary (5,6) and a single
star 3 are left. They are unbound. This can be indicated by writing it
as (1,2), (5,6), 3.

In the transition from the previous phase to the current one, at $\tau
= 0.0$, the internal structure is changed from \{[(2,3),1],(5,6)\} to
three unbound components, which are two binaries (1,2) and (5,6) and
one single star 3. This takes place in the following way.  Star 3 is
replaced by star 1. This exchange interaction exerts a kick on star 3
and the new binary (1,2). This kick unbinds the three
components. Since it is clear how the quintuple system is destroyed,
we do not investigate these interactions in any more detail.

\section{The prehistory of the first run of Paper I}
\label{sec:prehistory}

Section \ref{sec:history} began just before the preexisting binary
(3,5) is destroyed, and the first binary (1,2) emerges from the
subsystem (1,2,3,4,5,6). What we show in the present section is that
the history of these stars extends much further back.  In particular,
in order to investigate how the subsystem (1,2,3,4,5,6) and its binary
(3,5) were formed, we now follow the orbits of the subsystem
components and its surroundings before $\tau = -6.75$.  We shall,
however, confine the discussion to a presentation of results, and will
generally not describe the use of binding energies and work functions
on which the interpretation depends.

Figure \ref{fig:ebmax} gives an overview of binary activity over a
long period culminating with the formation of the $10kT$ binary (1,2)
at $\tau = 0$.  It shows the time evolution of the maximum binding
energy between any pair of stars at each time. Before $\tau=-6.75$, at
least one hard binary, which consists of stars 2 and 4, is formed, and
its binding energy reaches $9kT$ at $\tau = -43$ (figure
\ref{fig:ebmax}). The binary (2,4) persists for a while, exchanging
its components with intruders several times. The binary consists of
stars 6 and 9 at $\tau = -25.8$. However, the binary (6,9) is much
softer than the original binary (2,4). By $\tau = -12$, there is no
longer any binary with more than $2kT$.  Curiously, the components of
the binary at $\tau = -43$ belong to the subsystem whose evolution we
followed in Section \ref{sec:history}, and one of them is also a
member of the ``final'' binary.

\begin{figure*}
 \begin{center}
   \includegraphics[scale=1.2]{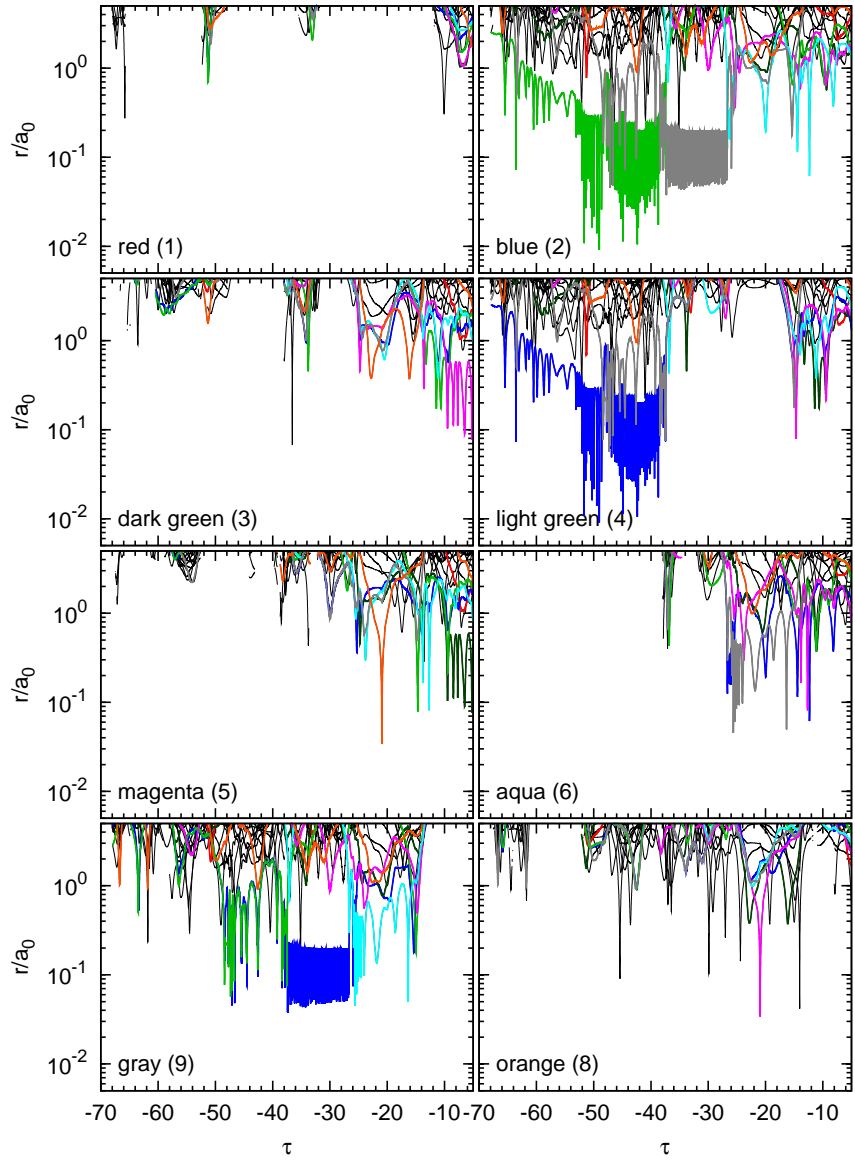}
 \end{center}
 \caption{The same as figure \ref{fig:paperI}, except that star 7 is
   replaced with star 9, shown in grey.}
 \label{fig:prehistory}
\end{figure*}

Now we check the formation and destruction processes of the $9kT$
binary in detail. Figure \ref{fig:prehistory} shows the distances
between the stars composing the subsystem mentioned in section
\ref{sec:history}.  However, star 7, which has no role in the
prehistory, is replaced with star 9.

We can see immediately that the $9kT$ binary consists of stars 2 and
4. By analysis analogous to that of Section \ref{sec:history}, it is
found that this binary is gradually hardened by encounters with
several distinct stars, including star 9, from $\tau = -66$
to $\tau = -52$. At $\tau = -48.5$, star 9 intrudes into the
binary. The binary and star 9 become a hierarchical triple system,
which survives until $\tau = -38$. At $\tau = -37$, star 4 is
exchanged with star 9, and a binary (2,9) is formed. The binary (2,9)
is almost unperturbed for a long time, until $\tau = -26.6$.

At $\tau = -26.6$, star 6 falls into the binary (2,9), and the three
stars 2, 6, and 9 form a temporary bound triple with a single ejection
of star 9 which returns at $\tau = -25.8$. The temporary bound triple
system ends up with the ejection of star 2, and the binary (6,9) is
formed. The binary (6,9) is only half as hard as its progenitor binary
(2,9) (as can just be seen in figure \ref{fig:ebmax}). The binary
(6,9) is perturbed by stars 5 ($\tau = -24$) and 8 ($\tau = -21$), and
finally destroyed by intrusions of stars 2, 3, 4, and 5 at $\tau =
-15$. We return to this event at the end of this section.

After that, there is no binary until a binary (3,4) appears at $\tau =
-12$. The binary component 4 is exchanged with star 5 at $\tau
=-9.5$. The binary (3,5) is the same as a binary which we see from
$\tau = -6.75$ to $-4.2$ in figure \ref{fig:paperI}. It was with that
binary that our discussion of the history began, in Section
\ref{sec:phase_1st}.

\begin{figure}
 \begin{center}
   \includegraphics[scale=1.8]{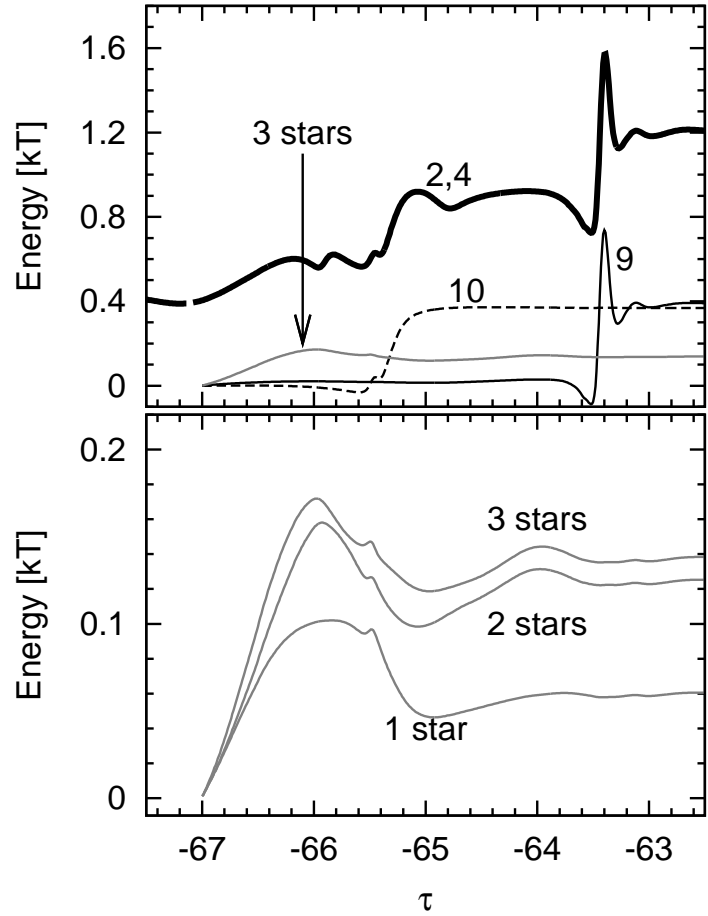}
 \end{center}
 \caption{Time evolution of the binding energy of a pair (2,4), and of
   work done on the pair. In the top panel, the binding energy is
   indicated by the thick black curve, the work done by stars 9 and 10
   by solid and dashed black curves, and the sum of work done by three
   stars by a solid gray curve. In the bottom panel, the work done by
   one, two and three of the three stars are indicated from bottom to
   top.}
 \label{fig:m1st_p1}
\end{figure}

\begin{figure}
 \begin{center}
   \includegraphics[scale=1.8]{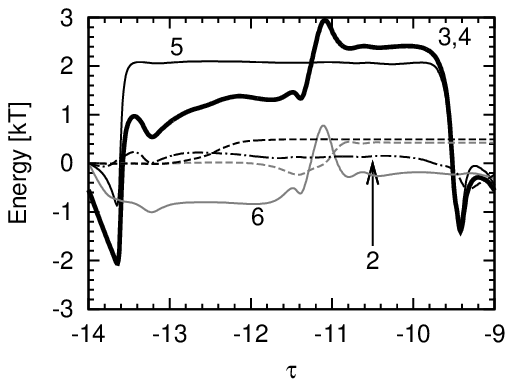}
 \end{center}
 \caption{Time evolution of the binding energy of a pair (3,4), and of
   work done on the pair. Solid black and gray curves indicate work
   done by stars 5 and 6, respectively, the dashed-dotted black curve
   indicates work done by star 2, and the other curves indicate work
   done by two other stars.}
 \label{fig:m1st_p3}
\end{figure}

In the prehistory, two binaries are formed not by exchange
interactions but by encounters involving more than three stars,
i.e. binaries (2,4) at $\tau = -66$, and (3,4) at $\tau = -12$.  We
analyse their formation using work functions. Figure \ref{fig:m1st_p1}
shows the time evolution of the binding energy of the binary (2,4),
and work functions for the binary. The binary (2,4) becomes harder in
three phases from $\tau = -66$ to $-63.5$.  Its hardening involves
five stars: one set of three stars, which we do not identify
otherwise, and successively stars 10 and 9.  Similarly we can see from
figure \ref{fig:m1st_p3} that the binary (3,4) is formed and hardened
through encounters with stars 5 and 6, though it is also hardened by
two other (unnumbered) stars. Although star 2 is a member of the
subsystem (2,3,4,5,6) existing at this time, star 2 does not
contribute to the formation of the binary (3,4).

Finally, we focus on the formation of the subsystem (2,3,4,5,6) at
$\tau = -15$. Star 6 is a component of the preexisting binary (6,9).
Stars 2, 3, 4, and 5 dissolve the binary (6,9) around $\tau = -15$,
and share out the binding energy of the binary (6,9). Ejection of star
9 also contributes to the binding energy of the subsystem (2,3,4,5,6).

\section{Discussion}

Paper I established a new framework for understanding the formation of
the first long-lived binary in an equal-mass $N$-body system at the
end of core collapse.  There it was shown that the standard paradigm
of formation in a three-body encounter between single stars was very
incomplete, in the sense that the encounters which form and harden a
binary often involve four or more stars.  In the present paper we
sharpen and clarify this new picture, showing the complete dynamical
history of the first long-lived binary in a system with $N = 1024$.

The system we have studied is the first model from Paper I, identified
there as ``seed 1''.  The short narrative on this model in Paper I
dealt with the genesis of three hard binaries, labelled there (by
their components) as (1,5), (2,3) and (1,2).  In the present paper we
have nothing to add to what was said in Paper I on the formation of
the last of these, but our more detailed analysis reveals the
following phenomena:
\begin{enumerate}
\item[1.] In Paper I it was stated that the binary (1,5) formed in a
  four-body encounter, but in the present paper it has been shown (see
  figure \ref{fig:diagram}, $\tau\simeq-1.6$) to have
  formed in an interaction between a bound 5-body system and an
  interloper which leads to (i) one escaper, (ii) the binary and (iii)
  three loosely-bound companions.
\item[2.] Paper I stated that the binary (2,3) emerged from a system
  of 5 stars, but we have now seen (see figure \ref{fig:diagram},
  $\tau\simeq-0.5$) that the event was an encounter between a
  temporarily bound three-body system and two interlopers, leading to
  the ejection of a soft binary from the triple system and the
  formation of a binary with a loosely bound companion.
\item[3.] What was described in Paper I is simply the end-game in a
  remarkably prolonged sequence of interactions, one of which led to
  the formation of a binary with a binding energy of $\simeq9.5kT$
  (where $3NkT/2$ is the initial kinetic energy of the entire $N$-body
  system).  This binary did not survive, and subsequently there were
  periods when the maximum binding energy of any binary in the system
  was less than $2kT$.  Nevertheless, one component of the
  $\simeq9.5kT$-binary was also a component of the ``final'' binary.
\end{enumerate}
In the following subsections we consider a number of general issues
which are raised by these observations.

\subsection{The role of democratic resonances}\label{sec:dem-res}

Democratic resonance is a frequent outcome of three-body interactions
involving a hard binary \citep{1993ApJ...403..256H}, and leads to a
triple system with a binding energy close to that of the binary.

Democratic resonances can be remarkably long-lived.
\citet{2007MNRAS.379L..21M} found that the distribution of lifetimes
(time to disruption) is approximately exponential, with an $e$-folding
time approximately $t_{\rm d} \simeq 250$, in units where all masses
are unity, $G=1$ and the internal energy of the triple system is $E =
-1$.  This translates to $t_{\rm d} \simeq 250 Gm^{5/2}\vert
E\vert^{-3/2}$. If, as suggested above, $E$ is close to the energy of
the original binary, the virial radius of the triple system is
approximately $R = 4.5a$, where $a$ is the semi-major axis of the
binary. Assuming that gravitational focusing acts, as in the hard
binary limit, we estimate the cross section for the approach of a
fourth body to within a distance $R$ to be approximately $\Sigma =
8\pi GmR/V^2$, where $V$ is the relative speed of the fourth body and
the triple system when far apart.  Therefore the probability, $P_4$,
of such a four-body encounter during the mean lifetime of the triple
system is approximately
\begin{equation}
  P_4 \simeq 9000\times2^{3/2}\pi n \frac{(Gm)^{1/2}a^{5/2}}{V},
\end{equation}
where $n$ is the number density of single stars.  The relatively
large numerical coefficient is a combination of those in the
expressions for $t_{\rm d}$, $R$ and $\Sigma$, the latter two arising from
the relatively large mass of the triple system.

Suppose the encounter takes place in a core with one-dimensional
velocity dispersion $\sigma_{\rm c}$ and central number-density
$n_{\rm c}$.  Then the conventional dynamical core radius is $r_{\rm
  c} = 3\sigma_{\rm c}/\sqrt{4\pi Gmn_{\rm c}}$, and we shall write
$N_{\rm c} = 4\pi n_{\rm c} r_{\rm c}^3/3$ to represent the number of
stars in the core.  Approximating $n = n_{\rm c}$ and $V =
\sqrt{3}\sigma_{\rm c}$ we find that the probability of a four-body
encounter during the mean lifetime of a temporary three-body system is
\begin{equation}
  P_4\simeq \frac{2.0\times10^4}{N_{\rm c}^2}\left(\frac{3m\sigma_{\rm
      c}^2/2}{E_{\rm b}}\right)^{5/2},
\end{equation}
where $E_{\rm b}$, the binding energy of the binary, has been
expressed in terms of the mean kinetic energy of stars in the core.
Clearly, this expression has to be interpreted appropriately if it
exceeds unity.

The above theory helps us to understand how typical the evolution
studied in this paper is.  Consider, for example, the triple system
(1,5,6) which survives from $\tau \simeq -0.8$ until $-0.5$ (figure
\ref{fig:diagram}).  In terms of the mean kinetic energy of all stars
its binding energy is $E_{\rm b} \simeq 10\times 3kT/2$ (the top panel
of figure \ref{fig:tuple_4th}), and in terms of the mean kinetic of
stars in the core the factor will be substantially less than 10.  We
see, therefore, that the probability of a four-body encounter during
the lifetime of a temporarily bound triple system is large, even for a
core with $N_{\rm c}$ as large as 10 (say).  Such an example shows
that the four-body behaviour seen in the system under study must occur
quite frequently.

\subsection{Bound few-body systems}

As we have seen, it is easy to understand the formation of temporarily
bound three-body systems. But the history discussed in the present
paper has examples of temporarily bound systems with larger numbers of
stars.  Notable examples are the five-body systems to be found in the
interval $\tau\simeq-15$ to $-13$ and in the interval from $-4.2$ to
$-1.6$, which we refer to as $V_1, V_2$, respectively.  Each of these
appears to form around a smaller existing bound subsystem. $V_1$ forms
from the triple (6,8,9), which itself formed from the hard binary
(6,9) in a democratic resonant interaction, while $V_2$ forms when the
hard binary (3,5) captures three other single stars almost
simultaneously.  Roughly speaking, both of these five-body systems can
be viewed as five-body analogues of a democratic resonance. In each
case the binding energy of the natal binary, which acts as a kind of
nucleus, leads to a temporarily bound system with five stars.

A complementary (but not contradictory) view of these five-body
systems is that they represent temporary fluctuations in the size of
the core. Such fluctuations are a notable feature of any $N$-body
simulation, but their dynamical origin is little understood. The
temporary capture of three single stars by a binary, which finally
results in energetic ejections and in a corresponding increase of the
binary's binding energy, may be one mechanism by which the core
contracts to small values of the core radius.  This is a rather more
dynamical and active picture of extreme fluctuations of the core,
compared to what is perhaps the more common understanding, i.e that
fluctuations are caused by the random phases of stars as they orbit in
and out of the core.

There is a sense in which the core of an $N$-body system is always a
bound subsystem.  If we compute the binding energy of that part of an
isothermal model lying inside radius $r$, we find that its value is
positive if $r > 1.58r_{\rm c}$ approximately.  In terms of star
numbers, the binding energy is positive if $N > 2.22N_{\rm c}$. In
other words, if we removed the stars outside this radius, the
remaining stars inside this radius would form a bound system which
could not disperse to infinity as single stars. From this point of
view one might think of a temporarily bound 5-body subsystem as the
core of the entire system at a time when its core radius is extremely
small.

In using the virial theorem, in the above discussion, we ignore
pressure exerted on an imaginary sphere around the core. This pressure
term is formally required in analysing the virial balance of a subset
of a system (here, the cluster core) in hydrostatic equilibrium. This
term, proportional to the stellar density at the surface of the
imaginary sphere, is negligible compared to the density in the
core. This implies that we can treat the core as a nearly isolated
system, as far as the use of the virial theorem is concerned.

\subsection{Immortal binaries}

The conventional view of core bounce is that high densities towards
the end of core collapse lead to the formation of a hard binary, which
interacts with other stars to heat the core and prevent its further
collapse.  In the process, the binary hardens almost relentlessly, and
it is usually assumed that the binary will not be destroyed (ionised)
in the stream of interactions. Instead, it eventually undergoes an
encounter so energetic that the binary itself is ejected, at least
from the core.  It is therefore perhaps a surprise to observe a binary
(the pair 2,4), with energy about $9.5kT$, which appears to be
essentially destroyed, leading to a period when there is no binary
with an energy above $2kT$.

The probability that a hard binary is destroyed has been considered in
quantitative detail by \citet{1993ApJ...403..271G}, in a theoretical
study based on the picture of binary formation and evolution in a
uniform background of single stars.  Their results (their fig. 2b)
imply that a $9.5kT$ binary has a disruption probability of only about
$0.5\%$, reinforcing the unexpected behaviour of the binary to which
we have drawn attention.  On the other hand we have argued that there
are episodes in the subsequent evolution of this binary (and its
offspring) in which the core is a compact few-body system, and in that
situation the mean kinetic energy in the ``core'' may considerably
exceed that in the entire system (which determines the value of $kT$).
In its environment, then, the binary is not as hard as the numerical
result suggests, and the probability of its disruption is much higher.
Indeed \citet{1993ApJ...403..271G} show that the probability rises to
$50\%$ for a binary of energy about $2.9kT$, and so the probability of
disruption in a compact core is certainly much enhanced.  Clearly the
limit of $10kT$, which was selected in Paper I as the end-point of the
analysis, is not robust, though in the case studied here, the energy
of the final binary is actually a more comfortable
$16kT$.

The reason why the $9.5kT$ binary is not very hard is due to our
definition of ``$kT$''. We define ``$kT$'' using the velocities of the
stars at the initial time, not at the current time, and these stars
are distributed throughout the whole cluster, and not only in the
cluster core (see Section~\ref{sec:summaryI}). When desired, all
binding energies can be simply converted to $N$-body units. Our
approach has the advantage that it avoids the difficulties of trying
to define time-dependent or space-dependent values of $kT$.

Here are two examples of these difficulties, in defining `co-moving'
values of $kT$. First, it is not clear when or even whether to use the
velocities of binary components, or only the centre-of-mass
velocities, when computing the ``$kT$'' value of a binary, when
energetic binaries are present. When such binaries are relatively
isolated, we could use their centre-of-mass velocities. However, in
cases of strong interactions with the environment, component
velocities might be more appropriate, especially during resonance
interactions. Second, the value of ``$kT$'' would behave
discontinuously, if we were to define ``$kT$'' strictly by velocities
of stars in a cluster core, each time membership of the core changed.

\subsection{Delayed Core Bounce}

There remains the question of why the traditional estimates of
conditions at core bounce are wrong.  As discussed in Paper I, and in
some more detail in Section 1 of the present paper, several treatments
in the nineteen eighties showed compelling arguments for core bounce
to occur when the size of the core had shrunk to contain a few dozen
stars.  These arguments were based on the idea that core bounce occurs
when two processes are balanced: the loss of heat from the core by
two-body relaxation, causing it to shrink, and the production of heat
by hard binaries formed in three-body encounters.  When the latter
exceeds the former, core collapse can be reversed.

In these estimates, the energy generation rate is calculated by
multiplying an estimate of the formation rate by an estimate of the
total energy emitted by a binary while it remains inside the cluster.
In reality, however, it takes time for this energy to be emitted, and
it could be argued that core bounce takes place provided that a hard
binary has formed and that it heats the environment fast enough.  At
the very least, core bounce cannot occur until the first hard binary
has formed, and it is quite possible that this leads to a different
condition for core bounce, one in which the number of stars in the
core is smaller.

Even such an estimate for the time of formation of the first hard
binary is likely to be a poor guide to the occurrence of core bounce:
as we saw in the previous subsection, the emergence of an effectively
immortal hard binary is surprisingly difficult.  This in itself
results in a delay in core bounce, which therefore takes place at a
smaller core star number.

We conclude that reliable heat production from a hard binary will make
itself felt only after the core has shrunk significantly further,
after the point where the core traditionally was estimated to contain
a few dozen stars.  The arguments presented above, though qualitative,
are consistent with a core dwindling to contain only half a dozen
stars, as observed in the simulations presented in Paper I.

\section{Conclusion}

We have dissected the spacetime history of the formation of the first
hard binary in a $1024$-body run, in microscopic detail. This paper
is the second in a series, following Paper I in which we presented the
first such microscopic observation of hard binary formation during
core collapse.  The single run that we have investigated here in great
detail is the very first run we presented in that paper.  The main
improvements over Paper I are:
\begin{enumerate}
\item[1.] We have introduced a new type of reaction diagram, somewhat
  similar to Feynman diagrams in perturbative quantum field theory
  calculations, and also similar to what was used by
  \citet{1983ApJ...268..319H} in the top part of their fig. 3 (section
  3.2).
\item[2.] We have introduced a new tool, in the form of work
  functions (section 3.1).
\item[3.] We have introduced another new tool, subcluster analysis
  (section 3.1).
\item[4.] We have highlighted the central role played by democratic
  resonances, especially three-body resonances whose longevity makes
  it likely that many-body interactions take place in the core of a
  star cluster around core bounce (section 3.7 and 5.1).
\item[5.] We have traced the network of reactions leading to the
  initial formation of the first hard binary back to earlier times,
  showing a complexity significantly larger even than what we had
  already unearthed in Paper I (section 4).
\item[6.] We have provided a new qualitative argument to derive the
  delay of core bounce, compared to standard expectations, based on
  the delay of hard binary heat production after formation (section
  5.4).
\end{enumerate}

It is interesting to note that we have employed four distinct levels
of analysis:
\begin{enumerate}
\item[1.] {\it Visual analysis} of the motions of stars in the core, using
an interactive visualisation tool in the form of an {\tt open-GL} program
(we have not stressed this initial phase, but it has helped to guide our
intuition and to resolve ambiguities).
\item[2.] {\it Geometric analysis} based on pairwise distances between
interacting stars.
\item[3.] {\it Energetic analysis} based on the binding energies of
pairs and higher-order multiple stars.
\item[4.] {\it Dynamic analysis} based on energy transfer between
tuples of stars.
\end{enumerate}
These steps lead to a compressed schematic rendition in the form of
the Feynman like diagram depicted in figure \ref{fig:diagram}.

The next step in our explorations will be to extend the applications
of our new techniques to a large number of $N$-body core collapse
simulations, for different values of $N$.  In order to do so, much of
the analysis presented here will have to be automated.  Ideally, all
of the figures presented here would be generated automatically by a
single analysis package.  In practice, the development of such a
package will remain a formidable challenge for quite a while to come.

A more modest step would be to develop improved tools to help generate
many of the figures semi-automatically, requiring far less time and
energy than has been the case for Paper I and the current paper, by
providing better graphics tools and other diagnostic tools covering
the physical properties of the core.

A next step could be to generate a kind of artificially intelligent
module that is trying to guess when an interesting network of
reactions starts and ends, around the time of core collapse, and
whether such a network includes the formation of a surviving hard
binary.  Using such a tool would still require human supervision to
check whether the results make sense, and to arbitrate in ambiguous
situations.

Ideally, after one or more steps, we could then build a software
system that fully automatically would produce all the diagrams
presented in this paper for any run, including the one introduced here
that resembles a Feynman diagram.

The results that we have presented could in principle be obtained from
an $N$-body simulation code like {\tt NBODY6}, which contains modules
that allow the user to output logs with information about binaries and
their hierarchy \citep{Aarseth01}. An analysis of these logs are
expected to produce the same result as we have obtained (for the same
numerical orbits), if the user has a way to deal with the huge amount
of data that would be produced. What we add here is a set of tools
that enable the user to analyse those kinds of data.

We could take a further step, and add more realistic effects to our
simulations, such as primordial binaries. It would be interesting to
elucidate whether or not the presence of hard primordial binaries
tends to suppress the formation of new binaries.

\section*{Acknowledgment}

Numerical simulations have been performed with computational
facilities at the Center for Computational Sciences in University of
Tsukuba. This work was supported in part by the FIRST project based on
the Grants-in-Aid for Specially Promoted Research by MEXT (16002003),
by KAKENHI(21244020), by Yukio Hayakawa Fund, by the Netherlands
Research Council NWO (grant \#643.200.503), and by the Netherlands
Research School for Astronomy (NOVA). Part of the work was done while
the authors visited the Center for Planetary Science (CPS) in Kobe,
Japan, during visits that were funded by the HPCI Strategic Program of
MEXT, and Lorentz Center in Leiden, the Netherlands. We are grateful
for their hospitality.

\bibliographystyle{elsarticle-harv}

\end{document}